\documentclass[a4paper,conference]{IEEEtran} 
\ifCLASSINFOpdf
\else
\fi
\usepackage[cmex10]{amsmath}
\usepackage{amssymb,verbatim}
\usepackage{graphicx}

\begin{document}
%
\title{%
Towards Quantum Enigma Cipher\\
-A  protocol for G bit/sec encryption based on 
discrimination property of non-orthogonal quantum states-
}

\author{
\IEEEauthorblockN{Osamu Hirota\\}
\IEEEauthorblockA{
Quantum ICT Research Institute, Tamagawa University\\
6-1-1 Tamagawa-gakuen, Machida, Tokyo 194-8610, Japan\\
{\footnotesize\tt E-mail: hirota@lab.tamagawa.ac.jp} \vspace*{-2.64ex}}
}

\maketitle

\begin{abstract}
This research note suggests a new way to realize  a high speed direct 
encryption based on quantum detection theory. 
The conventional cipher is designed by a mathematical algorithm and 
its security is evaluated by the complexity of the algorithm for 
 cryptanalysis and ability of computers. 
This kind of cipher cannot exceed the Shannon limit of cryptography, 
and it can be decrypted with probability one in principle by trying 
all the possible keys against the data length equal to the secret
 key length. 
A cipher with quantum effect in physical layer may 
exceed the Shannon limit of cryptography. 
The quantum stream cipher by $\alpha/\eta$ or Yuen-2000 protocol
 (Y-00) which  operates at Gbit/sec is a typical example of 
such a cipher. That is, ciphertext of mathematical cipher 
with a secret key is masked by quantum noise of laser light when 
an eavesdropper observes optical signals as a ciphertext of 
the mathematical cipher, while the legitimate receiver does not suffer 
the quantum noise effect. As a result, the inherent difference of accuracy of 
ciphertext between eavesdropper and legitimate receiver arises.
This is a necessary condition to exceed the Shannon limit of cryptography.
In this note, we present a new method to generate an inherent difference 
of accuracy of the ciphertext, taking into account a fundamental properties of 
quantum detection schemes.
\end{abstract}

%
\IEEEpeerreviewmaketitle
\section{Introduction}
 I introduced, in several public talks, the following situation:
[A new network scheme so called  ``Cloud computing system" based on data 
centers has recently attracted considerable attention. In that system, 
all data are communicated via a high speed optical network between 
a customer and data center or between data centers. 
There is a serious threat so called ``Eavesdropper data center business", 
which means the eavesdropper can get all data from the transmission line 
and sell specific data selected by the protocol analyzer to malicious 
people who want to get the secret data. This is a new business model 
of hacker in the era of cloud computing system.]
Surprisingly, it was exposed in 2013 by E.J.Snowden that US and UK governments 
 had collected all data from optical network. 
 
So one needs to consider `` Cyber attack against Layer-1 (physical layer)".
Technology of coupler for tapping has been 
developed by several institutes [1]. In addition, there are many optical 
monitor ports for network maintenance. Indeed physical layer
 of high speed data link is a defenseless. So, one can obtain the correct 
 ciphertext of mathematical cipher for paylord at Layer-2, and one can store 
 it at memory.
 
 This fact drove us
 to develop a special physical cipher for optical domain which protects
 ultra high speed data at Layer-1(physical layer). The requirements on 
the encryption system  are as follows:\\
\\
{\bf{Requirement of specifications:}}\\
 {{(1) Data-speed:}}1 Gbit/sec $\sim$ 100 Gbit/sec\\
 {{(2) Distance:}} 1000 Km $\sim$ 10000 Km \\
 {{(3) Encryption scheme:}} Symmetric Key Cipher \\
 {{(4) Security:}} Provable security, Secure against Brute force attack
(exhaustive search trial for secret key) by means of computer and also 
 physical devices. \\

So far, the standard encryption systems based on purely mathematical 
algorithm have been employed to ensure security of data and 
secret key. 
However, it is still difficult to quantitatively guarantee the security. 
Furthermore the eavesdropper can store the correct ciphertext from 
 communication lines as denoted at above.
  So one cannot deny the possibility of the 
decipherment of stored ciphertext by the discovery of 
the mathematical algorithm, or by the development of  high speed computers. 

The emerging development of physical cipher at the physical layer 
suggests a new way of building secure cloud computing system. 
A new concept of random cipher  based on quantum noise has been 
proposed [2]. It is called quantum noise randomized stream cipher 
or simply quantum stream cipher. The representative concrete protocol 
is  $\alpha/\eta$ or Y-00, and several implementation schemes have 
been realized [3,4,5]. 
The most important feature of this physical cipher is that the eavesdropper 
cannot get the correct ciphertext of mathematical cipher, for example
 a stream cipher by PRNG (pseudo randum number generator), from 
communication lines, while the legitimate 
user can get it based on a knowledge of secret key of PRNG. 
Thus, the ciphertext: $Y^B$ of the legitimate user and 
the ciphertext: $Y^E$ of the eavesdropper may be different as  
$Y^B \ne Y^E$. This can be enhanced by additional randomization 
such as Hirota-Kurosawa method [6]. 
Furthermore, the ciphertext of the eavesdropper
 becomes random by the real noise in her receiver.
The security is guaranteed by performance limitation  of 
the eavesdropper's receiver to get ciphertext and cryptanalysis 
ability to ciphertext.
A role of the Quantum Mechanics is to provide restrictions to accuracy of 
eavesdropper's ciphertext in physical layer, and to prevent, by 
physical law such as no-cloning theorem,  eavesdropper's  
attempt of the brute force attack.
In addition, there is a physical size limitation of devices for  
the physical brute force attack, which is also law of physics.

This type of scheme opens a new paradigm in the cryptology for physical layer. 
Indeed one can realize a cipher exceeding the Shannon limit of cryptography 
 so called  ``Quantum Enigma Cipher" that has been defined in our article[7].
However, one needs to investigate several randomization methods based on 
quantum mechanics.

 In the following sections, we will discuss a new method to 
realize a randomization of ciphertext based on quantum detection theory.

\section{Information theoretic basis for security of physical stream cipher}

It is well known that the Shannon limit of the symmetric key cipher 
derives a pessimistic theorem which justifies the one time pad. 
If the Shannon limit can be exceeded, one does not need hybrid cipher 
of QKD and one time pad. In this section, we give a survey of 
a new type of information theoretic basis on the symmetric key cipher.

In the conventional cipher, the ciphertext $Y$ is determined by 
the information bit $X$ and running key $K_R$ extended by PRNG with 
short initial key.
This is called non random cipher. On the other hand, one can introduce 
more general cipher system so called random cipher by classical noise 
such that the cipheretext is defined as follows:
\begin{equation}
Y_n=f(X_n,K_{R(n)}, r_n)
\end{equation}
where $r_n$ is noise. However, 
in the Shannon theory for non random or random  cipher, 
the information theoretic security on data is given
 as follows:\\
\\
{{\bf{Theorem 1:}}}(Shannon, 1949 [8])\\
The information theoretic security against ciphertext only attack 
on data has the following limit.
\begin{equation}
H(X|Y) \le H(K_s)
\end{equation}

This is called Shannon limit for the symmetric key cipher. 
In addition, the information theoretic security on key  is given by 
\begin{equation}
H(K_s|Y) \le H(K_s)
\end{equation}

Although the equality of Eq(2),(3) 
can be realized by ideal one time pad, there is no way to exceed 
the Shannon limit in the conventional encryption schemes.
To exceed  the Shannon limit is also essential for secure fresh key 
generation by communication or information theoretic security 
against known plaintext attack in the symmetric key cipher.

A random cipher by noise may provide a new category of the cipher, and 
so far many randomized stream ciphers have been proposed in the literature
 of cryptology.
However, at present, there does not exist an attractive cipher. 
The reason comes from the fact that the ciphertext of eavesdropper 
and that of legitimate user are the same in the classical system. 
 
It has been pointed out that there exists an attractive cipher, 
which gets out the frame of the Shannon theory of cryptology, 
by means of the combination of the concept of the private randomization
 of C.F.Gauss and the optical-quantum communication system [2].
 The crucial property of 
such a cipher is that the ciphertexts of eavesdropper and  
legitimate user may be different. Then it has a potential to exceed 
the Shannon limit, and may lead to realization of ``Quantum Enigma Cipher"[7].
In the following, we will give short review on the basis of cipher 
that exceeds the Shannon limit.
\\
\\
{{\bf{Theorem 2:}}} Let us denote the received ciphertexts of
 legitimate receiver and eavesdropper as follows:
\begin{eqnarray}
{Y_n}^B&=&\{{y_1}^B, {y_2}^B, {y_3}^B, \dots \}, \nonumber \\
{Y_n}^E&=&\{{y_1}^E, {y_2}^E, {y_3}^E, \dots \}
\end{eqnarray}
The necessary condition to exceed the Shannon limit is that 
the ciphertext of eavesdropper and that of legitimate user are different
 [for example, see 9]. 
\begin{equation}
{Y_n}^B=f(X_n,K_{R(n)}, {r_n}^B)  \ne  {Y_n}^E=f(X_n,K_{R(n)}, {r_n}^E)
\end{equation}
and
\begin{equation}
P_e(Y_n^E) >> P_e(Y_n^B)
\end{equation}
which is a condition of the performance on an error of detection 
of ciphertext for eavesdropper and legitimate receivers.
\\
Still the sufficient condition to exceed the Shannon limit is not 
strict, but if the following relation is retained, one can say 
that the cipher exceeds the Shannon limit.
\begin{equation}
H(X|Y^E,K_s) > H(X|Y^B, K_s) =0
\end{equation} 
$(Y^E,K_s) $ means that key is given after measurement.
The above equation means that eavesdropper cannot pin down 
the information bit even if she gets a secret key or unknown parameter 
for eavesdropper after her measurement of the ciphertext as physical signals.

\section{A new quantum stream cipher protocol based on controlled a priori 
probability}
According to the previous section, an important concept to exceed 
the Shannon limit is to realize the difference of accuracy of ciphertext 
between an eavesdropper's receiver and a legitimate receiver.
 In this section, we will propose a new way to realize it.

\subsection{Protocol}
A typical encryption scheme for physical layer in the network is 
a physical cipher
that scrambles transmission signals by integrating mathematical cipher 
with physical phenomena. So we employ a mathematical encryption box and a 
physical encryption box. This concept has a similarity to Triple DES with three different keys in classical cipher.  Let us show our method as follows:\\

(1) Here the data sequence is encrypted by PRNG with short secret key:$K_s$.
This part is an input to the physical encryption box:$Enc({\cal{P}})$.\\

(2) The ciphertext generated from PRNG is translated into  
$M$-ary  signals. \\

(3) An additional scheme as physical cipher is only to control  
a priori probability of $M$-ary signals by certain control box:$Enc({\cal{P}})$. 
The signals  from $Enc({\cal{P}})$ are mapped on  ``$M$-ary asymmetry 
non-orthogonal quantum states" at laser light transmitter, 
corresponding to one to one. ${\cal{P}}$ is a physical secret key 
of physical cipher. Thus,the legitimate receiver knows 
a priori probability, but the eavesdropper does not.\\

(4) The eavesdropper's receiver and the legitimate receiver will 
employ quantum optimum $M$-ary detection receiver. 
Since the legitimate receiver knows a priori probability, 
and the eavesdropper does not, the legitimate receiver can employ 
quantum Bayes strategy, but the eavesdropper has to employ 
quantum minimax strategy.\\

In general, the performance of quantum Bayes strategy is better than that of 
quantum minimax strategy. This means that error of the legitimate 
 receiver is smaller than that of the eavesdropper. 
 That is, we have\\
 
{{\bf{Theorem 3:}}}

 \begin{equation} 
 P_e(Bayes) \leq  P_e(minimax)
 \end{equation}
 By using this difference, the legitimate users can establish 
secure direct communication, because the  accuracy of the 
eavesdropper's ciphertext is deteriorated.
Let us give  such a principle in the following.

\subsection{Quantum detection theory}
Let us first describe the theory of quantum Bayes criterion. 
The optimum condition is given as follows:

The evaluation function of quantum Beys strategy is as follows:
\begin{equation}
\min\limits_{{\bf{\Pi}}}\sum_{i} \sum_{j} \xi_i C_{ji} 
Tr {\bf{\rho} }_i {\Pi}_i
\end{equation}
where, $ {\bf{\Pi}}=\{{\Pi}_j\}$ is POVM(positive operator valued measure).
As usual, we define the risk operator as follows:
\begin{equation}
{W}_j\equiv \sum^M_{j=1}\xi_i\it{C}_{ji}{\bf{\rho}}_i
\end{equation}
\begin{equation}
\Gamma=\sum^M_{j=1}{\Pi}_j{\bf{W}}_j=\sum^M_{j=1}{W}_j{\Pi}_j
\end{equation}

Here, we assume that  $ {C}_{ji}=1\ (i\ne j)$C$ {C}_{ji}=0\ (i=j)$. 
Then the criterion becomes average error probability. 
\begin{equation}
\min\limits_{\bf{\Pi}}P_e=\min\limits_{\bf{\Pi}}
(1-\sum_{i} \xi_i Tr {\bf{\rho} }_i {\Pi}_i)
\end{equation}\\

The optimum condition for quantum Bayes strategy with respect to 
POVM were derived by Holevo[10]and Yuen[11], respectively.\\

{\bf{Theorem 4:}}
Optimum conditions for POVM of quantum Bayes strategy 
are given as follows:\\
\begin{eqnarray}
&&({W}_j-\Gamma){\Pi}_j={\Pi}_j({W}_j-\Gamma)=0,
\quad \forall j \nonumber \\
&&{\Pi}_j({W}_i-{W}_j){\Pi}_i=0,
\quad \forall i,j \nonumber \\
&&{W}_j-\Gamma\ge 0,\quad \forall j
\end{eqnarray}
where 
\begin{equation}
{W}_j\equiv \sum^M_{j=1}\xi_i\it{C}_{ji}{\rho}_i
\end{equation}
\begin{equation}
\Gamma=\sum^M_{j=1}{\Pi}_j{W}_j=\sum^M_{j=1}{W}_j{\Pi}_j
\end{equation}

In general, when a priori probability is unknown, one cannot 
apply Bayes strategy, and minimax strategy may be employed.
In the following, we introduce a quantum strategy 
so called quantum minimax strategy which was formulated by 
Hirota-Ikehara in 1978.

In quantum case, the criterion of quantum minimax strategy is given by 
\begin{equation}
\bar{{C}}_m=\min\limits_{{\{{\Pi}_j}\}}\cdot\max\limits_{\{{\xi}_i\}}
\sum_{i=1}^{M}\sum_{j=1}^{M}\xi_i {C}_{ji} {T}r\rho_i {{\Pi}}_j
\end{equation}
where $ {C}_{ji}=1\ (i\ne j)$C$ {C}_{ji}=0\ (i=j)$. 
Then the criterion becomes
\begin{equation}
P_{em}=\min\limits_{{\{{\Pi}_j}\}}\cdot\max\limits_{\{{\xi}_i\}}
\left\{1-\sum_{i=1}^{M}
\xi_i {T}r\rho_i {{\Pi}}_j \right\}
\end{equation}

A primitive analysis was published in 1982 [12] 
as follows:\\

{\bf{Theorem 5:}} 
Let $\{\xi_i\}$ and $\{{\Pi}_j\}$ be a priori 
probability and POVM, respectively. Then we have 
\begin{equation}
\min\limits_{{\{{\Pi}_j}\}}\cdot\max\limits_{\{{\xi}_i\}}P_{e}=
\max\limits_{\{{\xi}_i\}}\cdot\min\limits_{{\{{\Pi}_j}\}}P_{e}
\end{equation}

{\bf{Theorem 6:}} The optimum conditions for POVM is given by
\begin{eqnarray}
&& {T}r\rho_i {{\Pi}}_i= {T}r\rho_j {{\Pi}}_j, \qquad \forall i,j
\nonumber \\
&&({W}_j- {\varGamma}) {{\Pi}}_j= {{\Pi}}_j( {{W}}_j- 
{\varGamma})=0,  \qquad \forall j \nonumber \\
&&{\Pi}_j({W}_i-{W}_j){\Pi}_i=0,\qquad \forall i,j \nonumber \\
&& {{W}}_j- {\varGamma}\ge 0, \qquad \forall j
\end{eqnarray}

Recently, the mathematical progress for quantum minimax theory 
has been given by G.M.D'Ariano et al[13], K.Kato [14], F.Tanaka [15], and 
K.Nakahira et al [16]. 

Now we can easily show the proof of the theorem 3.
The space $\Omega$ of a priori probability is separable and 
the space $\cal{D}$ of POVM is compact. Hence this strategy problem 
as the zero sum two person game becomes strictly determined
 from Wald's theorem. So, the minimax value  given by such two person 
game is the quantum Bayes value with the worst a priori probability. 
The minimax value  is always larger than the quantum Bayes value.

Thus, a knowledge of a priori probability can be employed as an encryption 
key in physical layer.

\section{Concrete properties of difference between Bayes and minimax}
\subsection{Design of quantum state signals}
Let $\{\rho_j\}$ be an ensemble of quantum states and 
whole element states are non-orthogonal each other of a set of states.
In addition, let us assume that a geometrical structure is an asymmetric.\\

{\bf{Definition 1:}} 
Symmetric quantum state is defined by
\begin{equation}
\rho_j =(U^{\dagger})^{j-1} \rho_1(U)^{j-1},\quad j=1,2,3,\dots,M
\end{equation}
where $U$ is a certain unitary operator. The example is as follows:
Let $|\alpha>$ be coherent state. The symmetric three states are 
\begin{equation}
|\alpha \exp({i\pi/2})>,|\alpha \exp({i7\pi/6})>, 
|\alpha \exp({i11\pi/6})> 
\end{equation}
like three pointed star of Mercedes-Benz.\\

{\bf{Definition 2:}} 
Asymmetric quantum state is defined by
\begin{equation}
\rho_j \ne (U^{\dagger})^{j-1} \rho_1(U)^{j-1},\quad j=1,2,3,\dots,M
\end{equation}
An example of the asymmetric state is 
\begin{equation}
|\alpha \exp({i\pi/2})>,|\alpha \exp({i7\pi/6})>, 
|\alpha \exp({i(\theta + 11\pi/6})> 
\end{equation}
where $\theta \ne 0$.
Here we have open problem on the optimum condition to design 
an efficient model for real use as follows:
\begin{equation}
\eta=\max\limits_{{\cal{R}}}\max\limits_{{\cal{P}}} \frac{P_e(minimax)}
{P_e(Bayes)}
\end{equation}
where $\sigma^{(k)}=\{\rho_j^{(k)}\}:\quad j=1,2,3,\dots,M$,
${\cal{R}}=\{\sigma^{(k)}\}: \quad k=1,2,3,\dots$, 
$\Omega^{(k)}=\{\xi_j^{(k)}\}:\quad j=1,2,3,\dots,M$, and
${\cal{P}}=\{\Omega^{(k)}\}:\quad k=1,2,3,\dots$.

Although we do not have an answer on the above question, we can 
show some examples.

\subsection{Nakahira-Kato-Usuda iterative method for optimum solution}
In general, the calculation of optimization problem in quantum 
detection theory is very hard, when a system consisting of 
a priori probability and a set of quantum states is given. 
So the iterative methods have been proposed by many researchers [17,18].
Recently, Nakahira-Kato-Usuda have provided useful iterative method 
which is applicable to both quantum Bayes and quantum minimax strategy [19].
As an example, Nakahira demonstrated properties for coherent state 
PSK(phase shift keying) of $M$=3 [20]. He showed 
\begin{equation}
\frac{P_e(minimax)}{P_e(Bayes)} \sim 10
\end{equation}
where $<n>=|\alpha|^2=1$. Although one needs to clarify the case of 
$<n> =10^4 \sim 10^6$ at transmitter and $M$ $>>$1, 
the above example is useful for estimation 
of the performance of this protocol. More realistic case will 
be reported in the future articles.

\subsection{Conjecture of good performance}
The eavesdropper can make a tapping at close the transmitter, so one 
needs to take into account the transmission loss for the performance 
advantage of the legitimate receiver. The quadrature amplitude 
modulation (QAM) is the most efficient modulation scheme under the 
average power constraint [21]. Keeping the good error performance of 
the legitimate receiver, one can design the QAM signal constellation 
such that the error performance of the eavesdropper's receiver becomes 
the worst. We will report the concrete design in the next paper.

\section{Security of total system}
Quantum enigma cipher which is the generalized 
symmetric key cipher as a physical cipher is designed by utilizing 
the features of mathematical encryption and physical phenomena in 
transmission line. The task of the eavesdropper is to find a secret key 
of mathematical encryption box from ciphertext as physical signal.
However, the eavesdropper cannot get the correct ciphertext. So it prevends 
the mathematical analysis to search the secret key and the Brute forc attack 
(exhaustive search trial for secret key ) by computer.
Although the eavesdropper may employ a physical Brute force attack, 
she cannot realize it because of the law of physics 
such as several quantum no-go theorems or physical size limitation 
for devices to try $2^{|K_S|}$ (candidate of the correct key),
 where $|K_S|$ is key length of secret key 
of mathematical encryption box.
Thus, one can realize a cipher for physical layer 
such that the security of the total system does not depend on 
the ability of computers and mathematical algorithm.

Here let us estimate a performance of the quantitative security. It is usual 
in the cryptology to use  guessing probability when the security does not 
depend on computational ablility.\\
(1) QKD(with initial secret key)+One Time Pad:\\
Let the generated key length $|K_G|$ by QKD and the data length of one time pad
 be both $10^4$ bits. The guessing probability of the key sequence of $10^4$ 
 for one time pad in the real case is given by 
 \begin{eqnarray}
 P(K_G|Y)&\le& \frac{1}{2^{|K_G|}} + d \nonumber \\
 &\sim& 10^{-15}
 \end{eqnarray}
 where $d$ is the trace distance in QKD model.
 The data transmission speed is limited by the speed of QKD.\\
 (2) Quantum Enigma Cipher:\\
Let the secret key length $|K_s|$ of mathematical encryption box 
and the data length be 256 bits and $10^4$ $<<$ $|X|$ $\le 10^{|K_s|}
 \sim 10^{70}$ bits, 
resepctively.
 The guessing probability of the secret key in the optimum design may 
 be expected as follows:
 \begin{equation}
 P(K_s|Y) \sim 10^{-70}
 \end{equation}
 The data transmission speed is limited by the speed of mathematical encryption  and the speed of the conventional optical communication system.
  Thus, the quantum enigma cipher has a potential for providing the ultimate 
 encryption scheme in high speed optical data link.
 
\section{Conclusion}
Y-00 type quantum stream cipher as a typical example 
of physical cipher in physical layer already has been realized and 
has been applied to 1 Gbit/sec $\sim$ 100 Gbit/sec real optical transmission 
system [22][23].
Now, we are concerned with a scheme with the ultimate security 
such as Quantum Enigma Cipher. There are two ways. One is to generalize 
Y-00 type stream cipher by additional physical randomizations, and other
 is to employ a new protocol.
In this note, to stimulate a development of quantum enigma cipher, 
we have introduced a new protocol that is useful 
for constructing it. More detailed story including a compound system of 
Y-00 type and this method 
will be introduced in the conference on Quantum Physics and Nuclear 
Engineering held in London at March 2016.

\section*{Acknowledgment}
I am grateful to K.Nakahira, K.Kato, T.Usuda, and F.Futami 
for fruitful discussions, to Mercedes-Benz club and Ministry of Defense.



\end{document}